\documentclass[a4paper,12pt]{article}
\usepackage{epsfig}

\def\dofigs#1#2#3{\centerline{\epsfxsize=#1\epsfig{file=#2, width=7.5cm, 
height=7.5cm, angle=-90}%
\hfil\epsfxsize=#1\epsfig{file=#3,  width=7.5cm, height=7.5cm, angle=-90}}}

\newcommand{\be}{\begin{equation}}
\newcommand{\ee}{\end{equation}}

\newcommand{\bea}{\begin{eqnarray}}
\newcommand{\eea}{\end{eqnarray}}

\newcommand{\gsim}{\ \rlap{\raise 2pt\hbox{$>$}}{\lower 2pt \hbox{$\sim$}}\ }
\newcommand{\lsim}{\ \rlap{\raise 2pt\hbox{$<$}}{\lower 2pt \hbox{$\sim$}}\ }

\def\gm{$\tilde g$MSB~}
\def\mam{mAMSB~}
\def\gaam{$\tilde g$AMSB~}

\def\tgb{\tan{\beta}}

\def\bsg{$b\to s\gamma$~}
\def\gmu{(g--2)$_{\mu}$~}

\begin{document}
\begin{titlepage}
\begin{center}
August 2001
\hfill
HIP-2001-45/TH \\
\hfill
TIFR/TH/01-34

\vspace*{3cm}

{\large \bf Vacuum stability bounds 
in Anomaly and Gaugino Mediated
SUSY breaking models}

\vskip .4in
Emidio Gabrielli$^{a,}$\footnote{emidio.gabrielli@helsinki.fi},
Katri Huitu$^{a,}$\footnote{katri.huitu@helsinki.fi},
Sourov Roy$^{b,}$\footnote{sourov@theory.tifr.res.in}\\[.15in]
{\em $^a$Helsinki Institute of Physics,\\
     POB 64,00014 University of Helsinki, Finland\\[.15in]
$^b$Department of Theoretical Physics\\
Tata Institute of Fundamental Research\\ 
Homi Bhabha Road, Mumbai - 400 005, India}
     
\end{center}

\vskip .2in

\begin{abstract}

\end{abstract}
We constrain the parameter space of the minimal and 
gaugino--assisted anomaly mediation, and gaugino 
mediation models by requiring that 
the electroweak vacuum corresponds to the deepest minimum
of the scalar potential.
In the framework of anomaly mediation models we find strong 
lower bounds on slepton and squark masses.
In the gaugino mediation models the mass spectrum is forced to be
at the TeV scale.
We find extensive regions of the parameter space
which are ruled out, even at low $\tan{\beta}$.
The implications of these results on the g-2 of the muon are
also analyzed.
\end{titlepage}

\newpage

\noindent
Supersymmetry (SUSY) is considered to be one of the most probable
alternatives for the physics beyond the Standard Model.  
The most general version of the minimal supersymmetric standard model 
includes a large number of parameters with which the 
ignorance of the supersymmetry breaking is parametrized.
Unfortunately with all the parameters, the model becomes
untraceable, and consequently several ways to simplify the 
parameter space have been considered.
These simplifications are based on the fact that if it were known
how supersymmetry is broken, one could calculate the soft
supersymmetry breaking parameters.
There are a large number of supersymmetry breaking
scenarios, which give acceptable phenomenology at least for some
part of the parameter space.

In recent years the branes, which are typical in models with extra
dimensions, have been found to fit naturally with the idea of breaking  
supersymmetry in a hidden sector.
Inspired by extra dimensions, anomaly mediated (AMSB) \cite{anomaly}
and gaugino mediated (\gm) \cite{gaugino} supersymmetry 
breaking have been constructed.
Here we will assume two parallel branes, 
which are located in one extra dimension,
as proposed by Randall and Sundrum \cite{anomaly}.
One of the branes contains the hidden sector, while the other
brane contains the ordinary matter.
Gravity is in the bulk.
Since there are no tree-level couplings between the fields in the observable
and hidden sectors,
the anomaly mediated contribution may be the dominant one.
In the pure AMSB scenario the slepton masses have negative squares,
but there are several proposals to fix the scenario \cite{anomalyfix}.
The most straightforward way would be to add a constant term to
the scalar masses (minimal anomaly mediation, \mam).
With extra
dimensions assuming gauge multiplets in the higher dimensional bulk, 
it was
found in \cite{Kaplan} that at one loop the squared slepton masses 
obtain  contributions, which would be of the correct size for solving 
the slepton mass problem (gaugino assisted anomaly mediation,
\gaam).
Also in the gaugino mediation models,
the gauge superfields propagate in the bulk in addition to 
gravity, but in the $\tilde g$MSB models they couple at tree-level 
to a singlet at the SUSY breaking brane.
The gaugino becomes massive, if the vacuum expectation value (VEV) 
of the singlet is nonvanishing, thus breaking supersymmetry.

Since none of the ways to break supersymmetry is compelling,
it is essential to study all the ways to restrict the parameter
space of the different models.
In addition to experimental bounds, there are theoretical
requirements, which must be fulfilled.
In this letter we will study the unbounded directions of
the vacua of the SUSY breaking models described above.

The presence of a large number of charged and colored scalar 
fields can generate dangerous minima in the scalar effective 
potential ($V$) of MSSM giving rise to an unacceptable color and 
electric charge breaking \cite{clm}. 
In the analysis of vacuum stability bounds,
radiative corrections to $V$ play an important role.
They are necessary in order to stabilize $V$
under variations of the renormalization scale, since the exact
expression for the renormalized effective potential should not depend upon 
this scale.
Due to this property, the search for these minima can
be strongly simplified by choosing an appropriate scale ($\hat{Q}$)
for which the one-loop corrections are minimized. A useful
approximation consists of analyzing 
the minima of the tree-level potential ($V_{tree}$) evaluated at this scale
$\hat{Q}$ and then requiring that the dangerous minima are never
deeper than the real minimum.

A complete study of vacuum stability bounds
in MSSM has been recently carried out in \cite{clm}.
In this work, two classes of necessary and sufficient conditions 
have been found. 
The strongest ones come from avoiding directions in the field
space along which $V_{tree}$ (calculated at $\hat{Q}$ scale)
becomes unbounded from below (UFB). These are given by a set of three
conditions, namely UFB-1,2,3.
The other class of constraints follow from avoiding the  
charge and color breaking (CCB) minima deeper than the realistic 
minimum.\footnote{One should note, however, that
the local minima may have lifetime longer than the present age of
the Universe \cite{falsev}. 
In this case the unstable directions may be acceptable.}
However, even though these conditions have been obtained 
in a model independent way, the phenomenological analysis 
has been carried out in specific models
\cite{clm,ufbprev}.

In a more recent paper \cite{dks}, the vacuum stability bounds have been 
analyzed for the \mam models.
In this work it was found that
UFB constraints (UFBc) would set quite strong bounds on 
sparticle spectra. In particular, selectron mass below $380$ GeV and stau mass
below $270$ GeV can be ruled out by applying the present 
experimental bounds on sparticle masses, 
especially the lower bound on chargino mass from LEP2 \cite{lep2}.
Since the only difference in the parameter space between the 
\gaam and the \mam model 
lies in the nonuniversality of the extra contribution 
to the scalar masses, 
one should expect that also for $\tilde g$AMSB models one gets strong
bounds. 
The major difference in the models comes from the weighted sum over
the quadratic Casimir for the matter scalar representations, leading
to nonuniversal contributions, whose relative strengths are given by 
\cite{Kaplan}
\bea
v=(Q,u,d,L,e,H_2,H_1)=(\frac{21}{10},\frac{8}{5},\frac{7}{5},\frac{9}{10},
\frac{3}{5},\frac{9}{10},\frac{9}{10}).
\label{casimir}
\eea

If there is a tree-level coupling of gauge fields to the SUSY breaking
brane, with a singlet which receives a VEV, the gauginos get a SUSY
breaking mass \cite{gaugino}.
Minimally the $\tilde g$MSB model has three parameters,
namely the Higgs mixing parameter $\mu$, the common gaugino mass
$M_{1/2}$ and the compactification scale $M_c$.
Following Schmaltz and Skiba in \cite{gaugino}, we assume that at 
the compactification scale the soft breaking
$A$ parameters, as well as the soft scalar masses, vanish
and $M_c$ is in the range 
$M_{GUT}\lsim M_c\lsim M_{Planck}/10$.
However, since we are interested in analyzing a more general scenario than in 
\cite{gaugino}, we will take $\tgb$  as a free parameter.
This is effected by relaxing the condition 
of vanishing soft $B$ parameter at $M_c$ scale assumed in \cite{gaugino}.

Before presenting our results, we briefly recall the definition of
the strongest UFB condition, namely UFB-3 
in the notation of reference \cite{clm}.\footnote{
However, in our analysis 
the full set of UFBc in \cite{clm} has been 
taken into account. }
This is obtained by avoiding dangerous UFB directions along the
down type Higgs VEV $\langle H_1\rangle=0$, after
suitable choices for down--squarks and slepton VEVs have been taken in order
to cancel (or keep under control) the $SU(3)$, $SU(2)_L$, and $U(1)_Y$
D-terms. For any value of $H_2$, such that $|H_2| < M_X$ 
(where $M_X$ is the high scale where soft breaking terms are generated), 
and
\be
|H_2| > \sqrt{\frac{\mu^2}{4\lambda_{e_j}}+\frac{4m_{L_i}^2}{g_Y^2+g_2^2}}
-\frac{|\mu|}{2\lambda_{e_j}}
\label{cond_a}
\ee
the following UFB-3 (strongest) condition must be satisfied 
\be
V_{UFB-3}(Q=\hat{Q}) > V_{{\rm min}}(Q=M_S)
\label{UFB3}
\ee
where 
\be
V_{UFB-3}=\left(m_{H_2}^2 +m_{L_i}^2\right)|H_2|^2+
\frac{|\mu|}{\lambda_{e_j}}\left(m_{L_j}^2+
m_{e_j}^2+m_{L_i}^2\right)
|H_2|-\frac{2m_{L_i}^4}{\left(g_Y^2+g_2^2\right)}
\label{ufb3_a}
\ee
with $i\neq j$, where $\lambda_{e_i}$ are the Yukawa couplings of 
the leptons $e_i$. 
In Eq.(\ref{ufb3_a}), the $\mu$ term is the usual bilinear coupling
between the Higgs doublets $H_{1,2}$ in the superpotential, and 
$m^2_{H_{1,2}}$, $m^2_{L_i}$, and $m^2_{e_j}$ are the soft--breaking terms for 
the Higgs doublets, left--, and right--handed sleptons,
respectively (with $i,j=1,2,3$ generation indices). 
Finally, $g_Y$ and $g_2$ correspond to the hypercharge and weak gauge
couplings respectively.
If Eq.(\ref{cond_a}) is not satisfied, then the expression for 
$V_{UFB-3}$ is the following
\be
V_{UFB-3}=m_{H_2}^2|H_2|^2+
\frac{|\mu|}{\lambda_{e_j}}\left(m_{L_j}^2+
m_{e_j}^2\right)
|H_2|+\frac{1}{8} \left(g_Y^2+g_2^2\right)\left(|H_2|^2+
\frac{|\mu|}{\lambda_{e_j}}|H_2|\right)^2
\label{ufb3_b}
\ee
It is clear that the optimum choice for the 
more restrictive condition is obtained by the replacement of
$\lambda_{e_j}$ with 
the Yukawa coupling of the tau ($\lambda_{e_j}\rightarrow \lambda_{e_{3}}$)
in Eqs.(\ref{cond_a},\ref{ufb3_a},\ref{ufb3_b}).
In Eq.(\ref{UFB3}) $V_{{\rm min}}(\hat{Q}=M_S)$ is the value of the 
tree-level (neutral) scalar potential calculated at the realistic minimum
\be
V_{{\rm min}}=-\frac{1}{2\left(g_Y^2+g_2^2\right)}\left(
\sqrt{\left(m_{H_1}^2+m_{H_2}^2+2\mu^2\right)^2-4|m_3|^4}
-m_{H_1}^2+m_{H_2}^2\right)^2
\ee
and at the scale $M_S\simeq \sqrt{m_{\tilde{t}_L}m_{\tilde{t}_R}}$.
The appropriate scale $\hat{Q}$ where $V_{UFB-3}$ 
should be evaluated (in order to minimize 
the one-loop corrections) is given by \cite{clm}
\be
\hat{Q}\simeq Max(g_2|e|,\lambda_{top}|H_2|,g_2|H_2|,g_2|L_i|,M_S)
\ee
where
\be
|L_i|^2=-\frac{4m_{L_i}^2}{g_Y^2+g_2^2}+\left(|H_2|^2+|e|^2\right)~~,~~
|e|=\sqrt{\frac{|\mu|}{\lambda_{e_i}}|H_2|} \, .
\ee

In Figs. (1a,b)--(2a,b) and (3a,b) we show our results
for the anomaly and gaugino mediation models respectively.
In the first class of models, in addition 
to $\tgb\equiv \frac{<H_2>}{<H_1>}$ and the $\mu$ term,
there are two more free parameters, the gravitino mass $m_{3/2}$ and $m_0$,
which sets the mass scale for the extra contribution $\Delta m^2(i)$ 
to the scalar masses (with $i=H_{1,2},Q,u,d,L,e$).
In the \mam model 
$\Delta m^2(i)$ is taken universal for all the scalars, namely 
$\Delta m^2(i)=m_0^2$.
The $\mu$ term, as usual, is fixed by requiring the correct electroweak
breaking condition, while the sign of $\mu$ ($sign(\mu)$) 
remains a free parameter in both models.

Our results correspond to $sign(\mu)$ for which the SUSY contribution
to the anomalous magnetic moment of the muon \gmu is always positive.
This choice is motivated by the recent measurement
of \gmu at BNL, where a $2.6 \sigma$ deviation
from the SM prediction has been reported \cite{bnl}. In particular, 
this deviation favors large and positive SUSY contributions to \gmu
which are achieved by large $\tgb$ values.
However, while the UFB-3 condition in Eq.(\ref{UFB3}) 
does not depend on $sign(\mu)$, this sign affects the physical 
spectrum for chargino, sleptons, and squarks and so
their corresponding UFB bounds.

In Fig.(1a) we show our results, in the $(m_{3/2},m_0)$ plane, 
for the allowed and disallowed regions by UFBc.
In addition, the regions ruled out by experimental 
lower bounds on chargino and stau masses,
respectively $m_{\chi} < 86$ GeV and $m_{\tilde \tau} < 82$ GeV 
\cite{lep2}, are also indicated.
Continuous and dashed curves correspond respectively to $\tgb=5$ and $40$.
Since SUSY contributions to \gmu and \bsg are enhanced by $\tgb$, we show 
them only for $\tgb=40$.
In particular, green and red shaded regions indicate, 
respectively, the allowed area
of BNL deviation on \gmu at 2.6 $\sigma$ level and the excluded one
by \bsg at 90\% of C.L..
In  Fig.(1b) we show the UFB allowed and disallowed areas
in the $(m_{\chi},m_{\tilde \tau})$ and $(m_{\chi},m_{\tilde t})$ planes,
where $m_{\tilde t}$ indicate the mass of the lightest stop.

From these results we see that UFBc can
set quite strong bounds on the relevant parameter space of 
the \mam model. For instance, the disallowed regions coming from 
the lower bound on stau mass are well inside the UFB disallowed area.
The combined effect of the UFBc and the experimental 
lower bound on chargino mass,
set a lower bound on $m_0$ of the order of 400 GeV and
slightly depending on $\tgb$. Moreover, from the results in Fig.(1b)
and for $\tgb=5~(40)$, they can set lower bounds of the order of 400 (300) GeV 
and 550 (530) GeV  for the stau and stop masses respectively.
For the opposite $sign(\mu)$ we get respectively 360 (300) GeV 
and 470 (510) GeV.
Our results are in agreement with the corresponding ones 
in reference \cite{dks}.

These results can be roughly understood as follows.
The UFB-3 condition in Eq.(\ref{UFB3}) 
might become very strong due to the presence of
the first term proportional to $m_{H_2}^2 |H_2|^2$
in Eqs.(\ref{ufb3_a}) and (\ref{ufb3_b}).
This term is negative (due to the fact that $m_{H_2}^2$ 
has to ensure the correct electroweak symmetry breaking),
leaving $V_{\rm UFB-3}$ very deep for large values of $|H_2|$.
However, the larger are the scalar masses (obtained by increasing $m_0$)
the larger is the term $m_{L_i}^2|H_2|^2$ and the other ones
proportional to $|H_2|$, leaving UFB-3 weaker.
The main dependence of UFB-3 bounds with $\tgb$ is due to
the second term in Eqs.(\ref{ufb3_a})--(\ref{ufb3_b}) 
which is proportional to the
inverse of the Yukawa coupling of the tau ($\lambda_{e_3}$).
By taking into account that the coefficient of $\lambda^{-1}_{e_3}$
is always positive and $\lambda^{-1}_{e_3}$ decreases with $\tgb$, 
the larger is $\tgb$, the smaller the second term becomes, and
thus makes UFB-3 stronger.
However, from Fig.(1b) we see that the behavior of UFBc 
with $\tgb$ is reversed in $(m_{\chi},m_{\tilde \tau})$ plane, 
while in the $(m_{\chi},m_{\tilde t})$ one it is
almost independent on $\tgb$.
This is due to the fact that the stau mass
is more sensitive to $\tgb$ than the stop one.

In Fig.(2a,b) we show the same results as in Fig.(1a,b), 
but for the \gaam model. The
$m_0$ parameter is defined here as $\Delta m^2 (L)\equiv m_0^2$, where
$(L)$ is the slepton doublet, while
the other contributions to the scalar masses $\Delta m^2 (i)$ 
are predicted in terms of the ratio of weight Casimirs in 
Eq.(\ref{casimir}). Note that the above considerations 
about the UFB-3 constraint hold for this model as well.
From the results in Fig.(1a,b) we see that the 
\gaam  model is slightly  disfavored by UFBc, when compared to
the \mam one.
The main reason is that in \gaam model the terms $\Delta m^2(i=Q,u,d)$ 
for squarks are larger than $\Delta m^2(i=L,e)$ for sleptons,
due to larger Casimir factors in Eq.(\ref{casimir}).
Indeed, as shown in \cite{clm,ufbprev}, 
models with small slepton masses and large
squark masses are disfavored by UFB-3. This is because
the larger is the stop mass, the more negative is $m_{H_2}^2$, 
due to larger renormalization group contribution. Smaller 
$m_{L_i}^2, ~m_{e_j}^2$, make the $V_{UFB-3}$ more negative, and thus 
the UFB-3 bound becomes stronger. 
Finally, combining in Fig.(2b) experimental bounds on chargino mass with 
UFBc, we obtain, for $\tgb=5~(40)$,
lower bounds of the order of 450 (360) GeV and 670 (690) GeV
for stau and stop masses respectively.
For the opposite $sign(\mu)$ we get respectively 430 (360) GeV 
and 620 (680) GeV.

In Fig.(3a) and (3b) we present the results
for the \gm model, with $SU(5)$ and $SO(10)$ unified groups respectively.
Here the free parameters, in addition to $\mu$ and $\tgb$, are 
the gaugino mass $M_{1/2}$ at GUT scale and $t_c=\log(M_c/M_{GUT})$.
Thus, the allowed range of $t_c$ is $0 < t_c < 4$.
Since in this model 
the scalar masses ($\tilde{m}$) and trilinear couplings ($A$) are
radiatively generated through the renormalization group running
from the compactification scale down to the GUT scale, 
they will be proportional respectively to 
$\tilde m^2 \propto M_{1/2}^2~\alpha_{GUT}~t_c$, and 
$A \propto M_{1/2}~\alpha_{GUT}~t_c$ times averaged Casimir factors, where 
$\alpha_{GUT}$ is the gauge coupling at GUT scale.
For the exact expressions we have
used the results of Ref. \cite{gaugino}.
In Fig.(3a,b) we show the allowed and disallowed regions
in the $(M,t_c)$ plane, with $M=M_{1/2},~m_{\tilde \tau}$.
In particular, regions
below the continuous and dashed curves, corresponding to $\tgb=5$ and 20 
respectively, are excluded.
We stress that the corresponding plots for  $M=m_{\chi}$ and $M=m_{\tilde t}$
almost overlap the $M=M_{1/2}$ and $m_{\tilde \tau}$ curves respectively.
{}From these results we see that 
the SUSY spectrum is above 1 TeV for most of the parameter space,
even for $\tgb=5$. Besides, for $\tgb > 20$, the allowed regions are
above the dashed lines, leading to much stronger bounds.
These results can be understood by noting that
scalar masses are proportional to gaugino mass $M_{1/2}$, being induced by  
radiative corrections, and are enhanced by $t_c$
which parametrizes the size of corrections from extra dimensions.
As explained above, increasing $M_{1/2}$ (and $t_c$) increases
the scalar masses, relaxing the UFB-3 constraint.
Note that at $\tgb=20$, regions below $t_c \lsim 2.6$ and $t_c \lsim 1.5$
are excluded by UFBc for $SU(5)$ and $SO(10)$ respectively.
Roughly the same results apply for the opposite $sign(\mu)$.

Now we discuss the impact of UFBc on the \gmu of the muon.
Recently the predictions for \gmu in the framework of anomaly mediation and
gaugino mediation models have been analyzed in Ref. \cite{egh}. 
However, in this work, UFBc were not taken into account.
From the results in Figs.(1a),(2a)
we see that in both anomaly mediation models
these constraints strongly affect \gmu, 
leaving this class of models disfavored for explaining 
the BNL deviation (at $2\sigma$ level), especially \gaam.
However, in the framework of \gm models, the effect of UFBc on \gmu 
is very strong. As shown in Figs.(3a,b), the \gmu allowed regions
are completely ruled out by UFBc.

{\it Summary.}
We have analyzed restrictions from UFB constraints on AMSB and 
{$\tilde g$}MSB models, which are closely connected to the ideas of
extra dimensions.
Our results show that the experimental limits on chargino mass puts
strong bounds on slepton and squark masses in AMSB models, and the
whole mass spectrum is very high for {$\tilde g$}MSB models, leading to
problems with naturalness requirement for most of the
parameter space.
If the experimental deviation in \gmu is due to supersymmetry, only
small windows in the parameter space of the AMSB models remain
open, and the {$\tilde g$}MSB models are ruled out.

\begin{figure}[tpb]
\dofigs{3.1in}{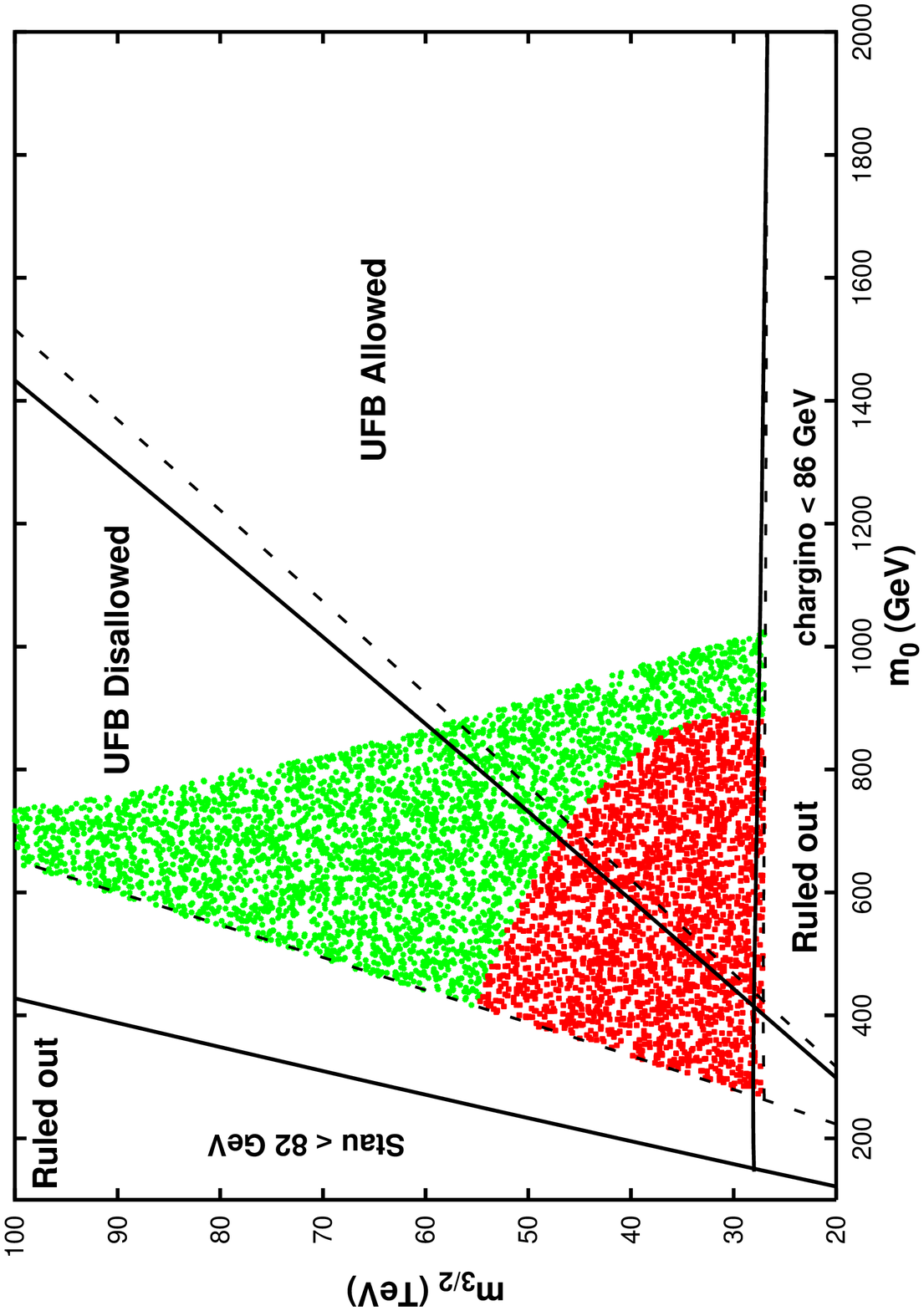}{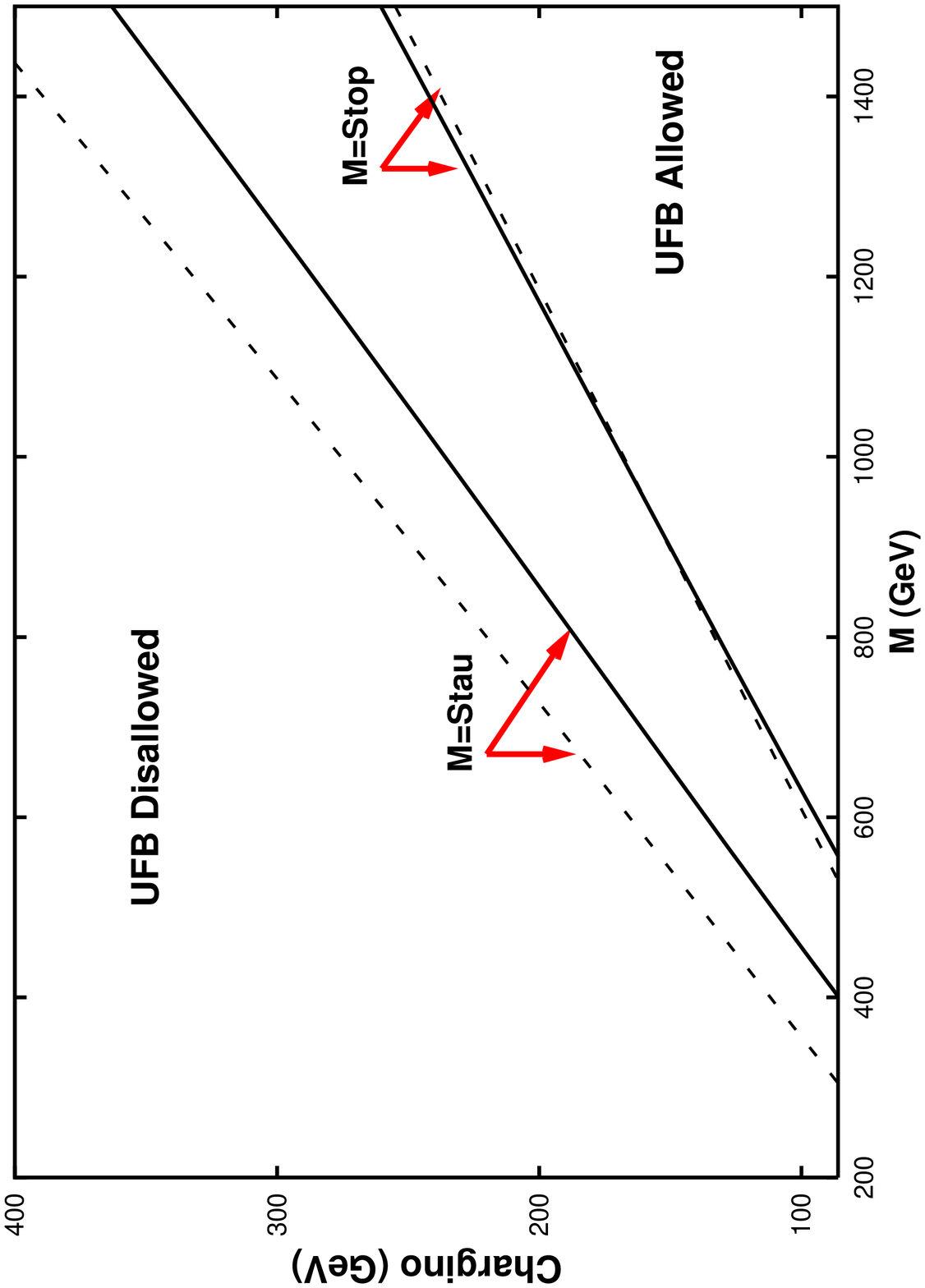}
\caption{\small On the left (a), the allowed region by 
UFB-constraints in 
($m_0,m_{3/2}$) plane is shown in the \mam model.  
Solid and dashed lines corresponds to $\tan\beta=5$ and 40
respectively.
For $\tan\beta$=40 we have also indicated the area forbidden by
$b\rightarrow s\gamma$ results (red shaded area) and allowed by
the \gmu results (green shaded area).
Other than UFB, lines come from experimental lower limits as indicated
in the plot.
On the right (b), the allowed ranges for $\tilde\tau$ and
$\tilde{t}$ masses are shown in ($m_{\chi},M$) plane.
Solid and dashed lines as before.}
\label{fig1}
\end{figure}

\begin{figure}[tpb]
\dofigs{3.1in}{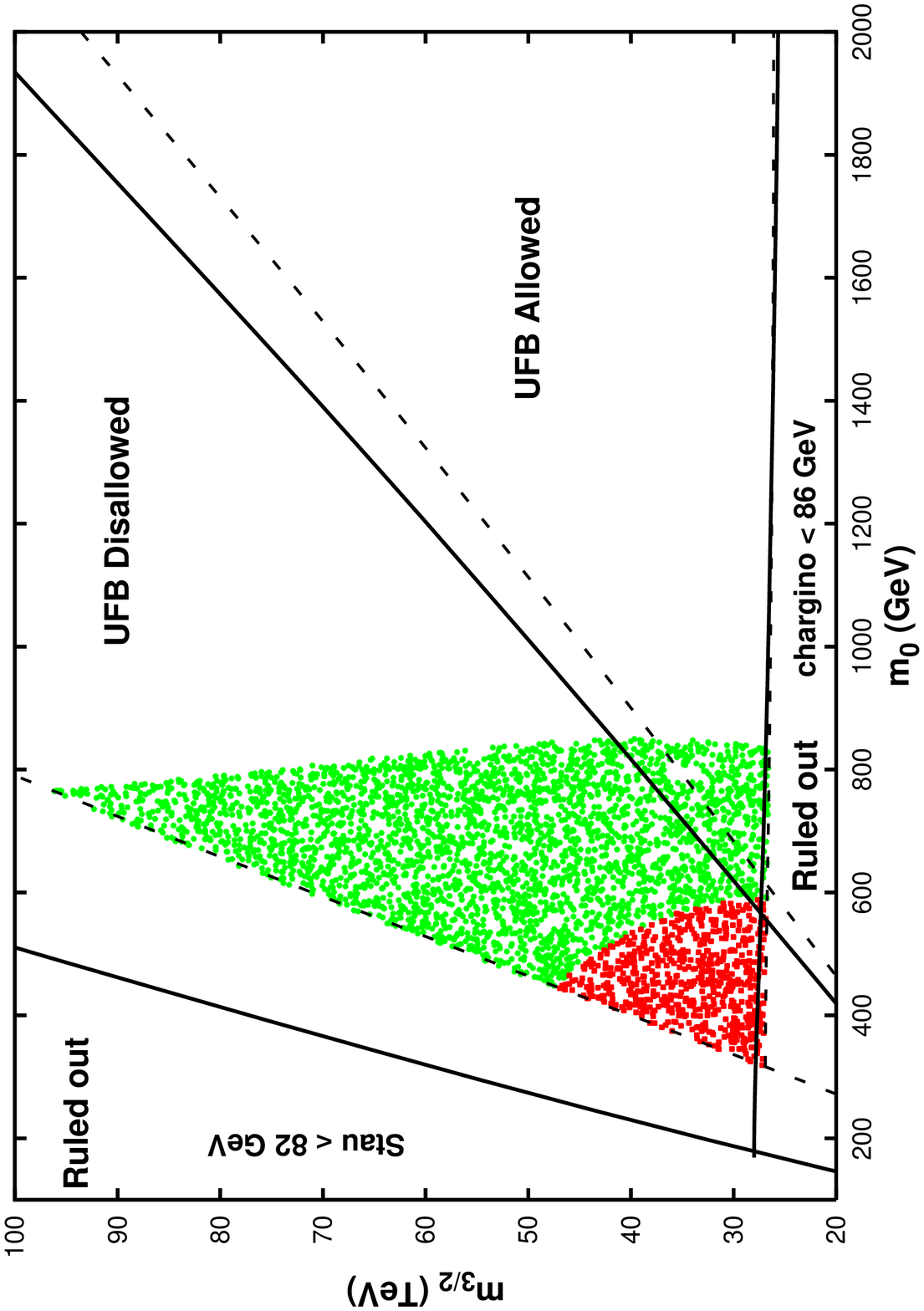}{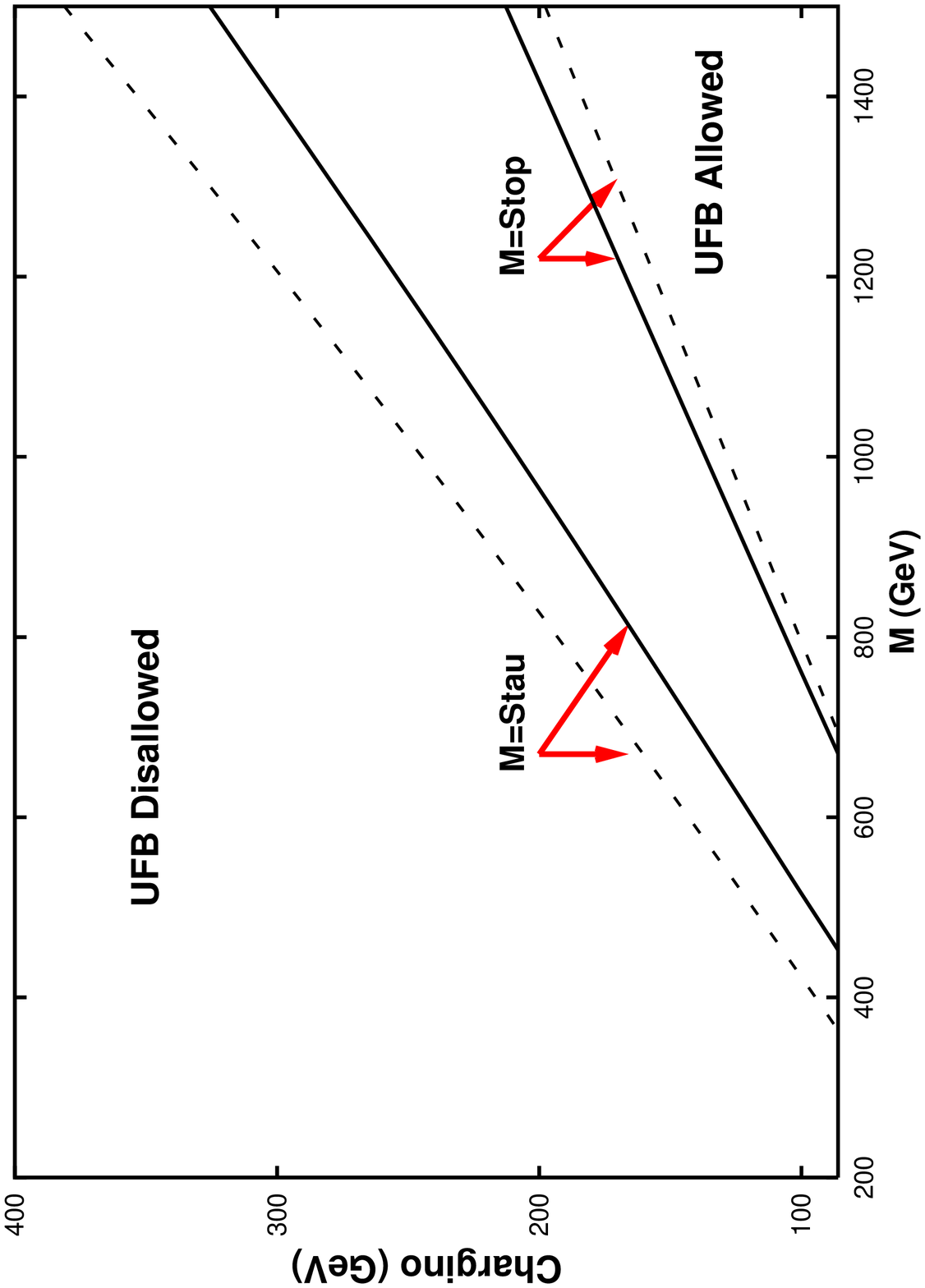}
\caption{{\small As Figure \ref{fig1} but for \gaam model.}}
\label{fig2}
\end{figure}

\begin{figure}[tpb]
\dofigs{3.1in}{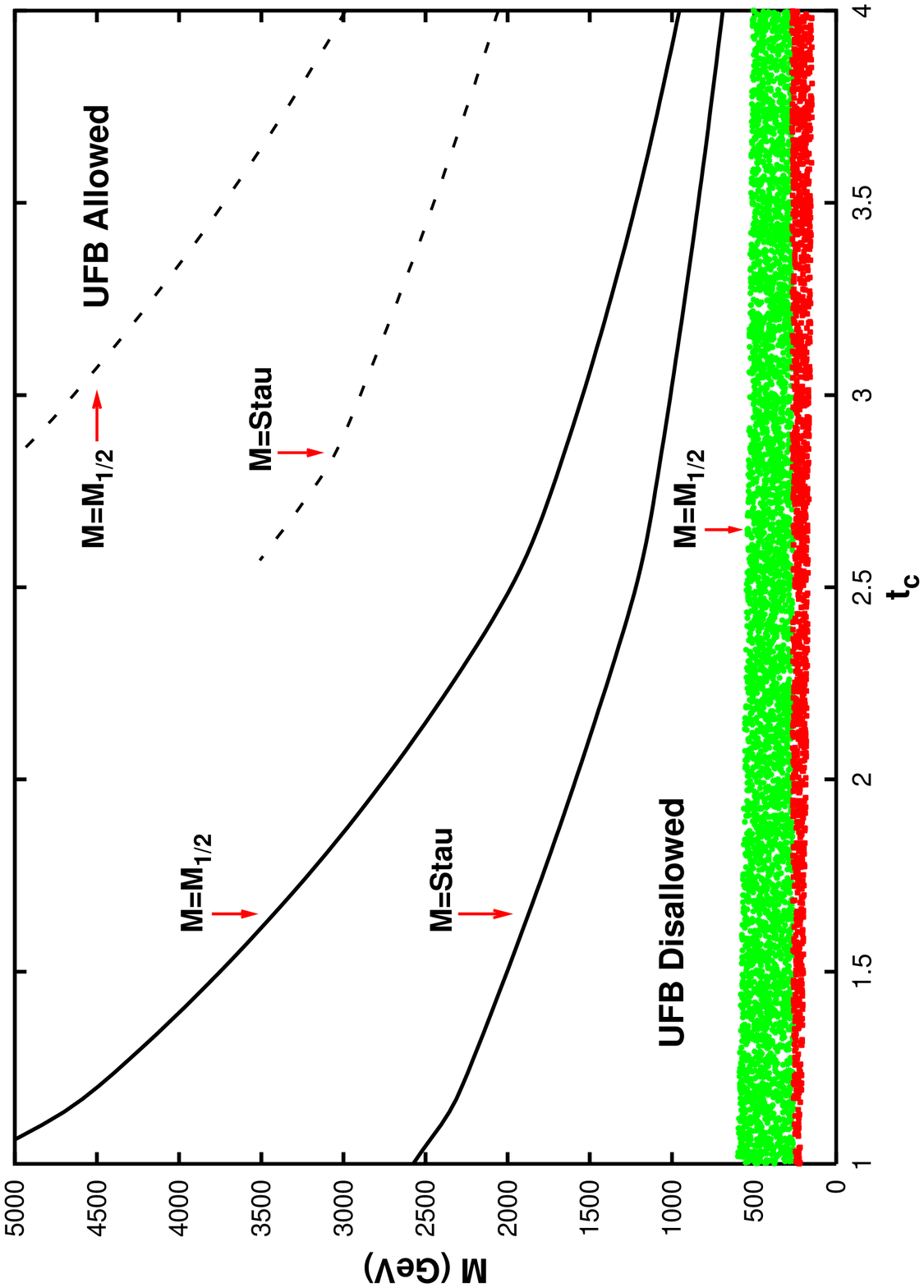}{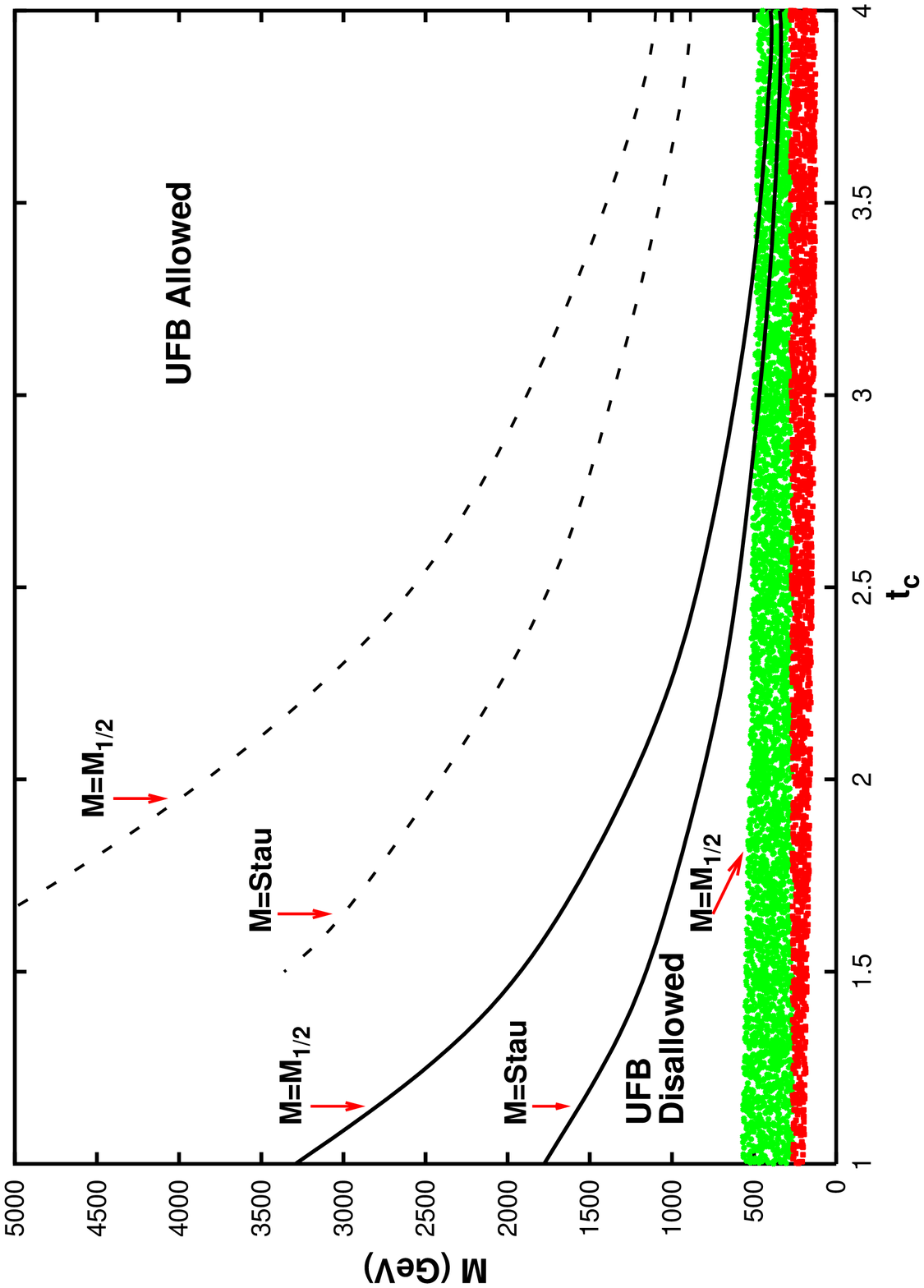}
\caption{{\small 
The allowed UFB ranges for the stau $(M=m_{\tilde\tau})$ and
gaugino $(M=M_{1/2})$ masses are shown in ($M,t_c$) plane for
the \gm model with $SU(5)$ (left (a)) and $SO(10)$ (right (b))
unified groups.
Solid and dashed lines correspond to $\tan\beta=5$ and 20
respectively.
Red and green shaded areas as in Fig.(\ref{fig1}), but for $\tgb=20$.
Regions below red shaded areas are excluded by $m_{\tilde \tau} < 82$ GeV.}}
\label{fig3}
\end{figure}

\vspace*{1cm}
\noindent
{\bf Acknowledgements}

\vspace*{0.5cm}
\noindent
This work was partially supported by the Academy of Finland
(project nos. 48787 and 163394).
S.R. wishes to acknowledge the hospitality provided by the 
Helsinki Institute of Physics, where a part of this work was done.


\begin{thebibliography}{100}


\bibitem{anomaly} L. Randall, R. Sundrum, B557, 79 (1999);
G.F. Giudice, M.A. Luty, H. Murayama, R. Rattazzi, 
JHEP 9812, 027 (1998) ; J.A. Bagger, T. Moroi, E. Poppitz, 
JHEP 0004, 009 (2000).

\bibitem{gaugino} D.E. Kaplan, G.D. Kribs, M. Schmaltz, 
Phys. Rev. D62, 035010 (2000);
Z.~Chacko, M.A. Luty, A.E. Nelson, E. Ponton, JHEP 0001, 003 (2000);
M.~Schmaltz, W. Skiba, Phys.Rev. D62, 095005 (2000).

\bibitem{anomalyfix} A. Pomarol, R. Rattazzi, JHEP 9905, 013 (1999);
R. Rattazzi, A. Strumia, J.D.~Wells, B576, 3 (2000);
Z.~Chacko, M. Luty, E. Pont\'{o}n, Y. Shadmi, Y. Shirman,
hep-ph/0006047;  E. Katz, Y. Shadmi, Y. Shirman, JHEP 9908, 015 (1999); 
Z. Chacko, M.A. Luty, I. Maksymsk, E. Pont\'{o}n,
JHEP 0004, 001 (2000); I. Jack, D.R.T. Jones, B482, 167 (2000);
M. Carena, K. Huitu, T. Kobayashi, B592, 164 (2001).

\bibitem{Kaplan} D.E. Kaplan, G.D. Kribs, JHEP 0009, 048 (2000).

\bibitem{clm} J.A. Casas, A. Lleyda, C. Mu\~{n}oz, Nucl. Phys. B471, 3 (1996).

\bibitem{ufbprev} 
A. Datta, A. Kundu, A. Samanta, Phys. Rev. D63, 015008 (2001);
S.A. Abel, B.C. Allanach, JHEP 0007, 037 (2000); 
J.A. Casas, A. Ibarra, C. Mu\~{n}oz, Nucl. Phys. B554, 67 (1999);
S. Abel, T. Falk, Phys.Lett. B444, 427 (1998);
J.A. Casas, S. Dimopoulos, Phys.Lett. B387, 107 (1996);
H. Baer, M. Brhlik, D. Castano, Phys.Rev. D54, 6944 (1996);
J.A. Casas, A. Lleyda, C. Mu\~{n}oz, Phys. Lett. B380, 59-67 (1996),
Phys. Lett. B389, 305 (1996).

\bibitem{falsev} A. Riotto, E. Roulet, Phys.Lett. B377, 60 (1996);
A. Kusenko, P. Langacker, G. Segre, Phys.Rev. D54, 5824 (1996).

\bibitem{dks} A. Datta, A. Kundu, A. Samanta, hep-ph/0101034.

\bibitem{lep2} M. Elsing (DELPHI Collaboration) presentations
on Feb. 27, 2001,  available
from http://www.cern.ch/\~{ }offline/physics\_links/lepc.html.

\bibitem{bnl} H.N. Brown et al. (Muon ($g-2$) Collaboration), 
Phys.Rev.Lett. 86, (2001) 2227; A. Czarnecki, W.J. Marciano, 
Phys. Rev. D64, (2001) 013014, and references therein.

\bibitem{egh} K. Enqvist, E. Gabrielli, K. Huitu, 
Phys. Lett. B512 (2001) 107.
\end{thebibliography}
\end{document}